# Optical spectroscopic investigation on the coupling of electronic and magnetic structure in multiferroic hexagonal $RMnO_3$ ($R$ = Gd, Tb, Dy, and Ho) thin films

Woo Seok Choi, Soon Jae Moon, Sung Seok A. Seo[a)], Daesu Lee, Jung Hyuk Lee[b)], Pattukkannu Murugavel[c)], and Tae Won Noh

*ReCOE & FPRD, Department of Physics and Astronomy, Seoul National University, Seoul 151-747, Korea*

Yun Sang Lee*

*Department of Physics, Soongsil University, Seoul 156-743, Korea*

We investigated the effects of temperature and magnetic field on the electronic structure of hexagonal $RMnO_3$ ($R$ = Gd, Tb, Dy, and Ho) thin films using optical spectroscopy. As the magnetic ordering of the system was disturbed, a systematic change in the electronic structure was commonly identified in this series. The optical absorption peak near 1.7 eV showed an unexpectedly large shift of more than 150 meV from 300 K to 15 K, accompanied by an anomaly of the shift at the Néel temperature. The magnetic field dependent measurement clearly revealed a sizable shift of the corresponding peak when a high magnetic field was applied. Our findings indicated strong coupling between the magnetic ordering and the electronic structure in the multiferroic hexagonal $RMnO_3$ compounds.

## I. INTRODUCTION

Recently, multiferroic oxides have attracted a great deal of attention due to their intrinsic coupling between magnetic and electric order parameters.[1-5] This magnetoelectric coupling within a single material could lead to novel applications as well as a new understanding of the underlying physics. One explanation proposed for the magnetoelectric coupling is the non-collinear antiferromagnetic spin structure. It has been suggested that such long-range magnetic ordering breaks the inversion symmetry in various ways,[3] such that ferroelectricity is eventually induced.[4, 6] Such intriguing magnetic exchange interactions strongly affect the electronic structure through the spin-charge interaction. Therefore, optical investigation of the electronic structure can provide indispensable information to improve our understanding of the microscopic mechanism of magnetoelectric coupling in multiferroic materials.

Hexagonal $R\mathrm{MnO_3}$ (Hexa-$R\mathrm{MnO_3}$, $R$: rare earth ions) is one of the typical multiferroic material with ferroelectric ordering typically occurring at temperatures above 590 K and antiferromagnetic ordering with a Néel temperature ($T_N$) between 56 K and 120 K.[7] Although the single crystals of hexa-$R\mathrm{MnO_3}$ compounds have been studied to some extent due to their interesting physical properties, the origin of its ferroelectricity is still a subject of debate.[4, 8] It has a layered structure consisting of alternating $\mathrm{MnO_5}$ triangular bipyramidal layers and rare earth layers in the out-of-plane direction. A key feature in this structure is a two-dimensional in-plane triangular lattice basis of magnetic $\mathrm{Mn^{3+}}$ ions.[4] Due to the geometrical frustration arising from the triangular basis and the resulting competition of magnetic interactions, the classical magnetic ground state of hexa-$R\mathrm{MnO_3}$ is characterized as a non-collinear spiral configuration in which the spins on each of the three sublattices are oriented at 120° from the other two.[3] In addition, the geometrical frustration in the magnetic ordering suppresses long-range Néel ordering, and prevents the magnetic materials with strong antiferromagnetic exchange interactions from aligning antiferromagnetically at their original Curie-Weiss temperature ($\theta_{CW}$).[9] The frustration parameter $f = |\theta_{CW}|/T_N$ in the hexa-$R\mathrm{MnO_3}$ system ranges from 5.8 to 10.3,[10, 11] indicating a high degree of magnetic frustration. Magnetoelectric coupling effects have been observed experimentally in the ferroelectric and antiferromagnetic phases of the hexa-$R\mathrm{MnO_3}$ system. Such effects include modification of the magnetic structure by an external electric field,[3] and variation of the static dielectric constant by an external magnetic field.[12]

Optical spectroscopic studies have previously indicated several intriguing features of hexa-$R\mathrm{MnO_3}$ in relation to strong spin-charge coupling. Souchkov *et al.* performed a temperature-dependent optical spectroscopic study on single crystal hexa-$\mathrm{LuMnO_3}$.[13] They observed a peak shift of the optical absorption structure at ~1.7 eV as a function of temperature and found an anomaly of the peak shift at $T_N$. They suggested that the temperature dependence of the absorption peak could be related to the antiferromagnetic exchange coupling in hexa-$LuMnO_3$. Later, Rai *et al.* reported reflectance spectra for hexa-$HoMnO_3$ single crystals.[14] Similar to the earlier study, they found anomalies in the dielectric constant at ~1.7 eV, indicating changes in the dielectric response with variation of temperature and/or magnetic field.

While previous studies have revealed intriguing optical responses in some bulk hexa-$R$MnO$_3$ samples, there have been few studies to investigate the effects of the type of $R$ ion on the electronic responses. To obtain an in-depth understanding of the intriguing optical properties in terms of the coupling between magnetic ordering and the electronic structure, systematic investigations of compounds in the hexa-$R$MnO$_3$ series with different $R$ ions are necessary. It should be noted that the compounds $GdMnO_3$, $TbMnO_3$, and $DyMnO_3$ have an orthorhombic structure in nature due to the relatively large size of the $R$ ions. However, they can be fabricated into thin films that contain a hexagonal phase with the epitaxial stabilization technique.[12] Using such novel thin film samples enables the number of compounds in the hexa-$R$MnO$_3$ series to be increased. From this extended hexagonal phase space, a better understanding on multiferroic properties can be obtained.

Here, we report a study of the effects of temperature and the magnetic field on the optical absorption features in multiferroic hexa-$R$MnO$_3$ (where $R$ = Gd, Tb, Dy, and Ho) thin films. We observed the changes in the electronic structure of the four compounds in the hexa-$R$MnO$_3$ series when the magnetic ordering of the system was disturbed by external parameters, such as temperature and/or high magnetic field. We found that the absorption structure near 1.7 eV showed an exceptionally large blue shift with decreasing temperature together with a maximum shift at around $T_N$. Such behavior is interpreted as being closely associated with the antiferromagnetic exchange coupling. Our findings from the optical spectroscopic experiments under a strong external magnetic field also imply a strong coupling between the magnetic ordering and the electronic structure in the multiferroic hexa-$R$MnO$_3$ compounds.

## II. EXPERIMENTAL METHODS

We fabricated high-quality hexa-$R$MnO$_3$ ($R$ = Gd, Tb, Dy, and Ho) thin films using pulsed laser deposition techniques. We used single crystalline yttria-stabilized zirconia (YSZ) (111) as the substrate, because the hexagonal in-plane net of the YSZ substrate stabilizes the hexagonal phase of $R$MnO$_3$ through good lattice matching between the substrate and thin films grown on the substrate. More details about the growth conditions and structural characterization of the thin films have been reported elsewhere.[12, 15-17] Transmittance optical measurements were performed in the photon energy range of 0.1–6.0 eV. We used an FT-IR spectrometer (Bruker IFS66v/S) and a grating-type spectrophotometer (CARY 5G) in the photon energy region of 0.1–1.2 eV and 0.4–6.0 eV, respectively. The optical absorption spectra were calculated from the transmittance spectra using the simple relation $\alpha(\omega) = [\log Tr(\omega) - \log Tr_{sub}(\omega)]/d$, where $\alpha(\omega)$ is the optical absorption, $Tr(\omega)$ is the transmittance of the thin film on the substrate, $Tr_{sub}(\omega)$ is the transmittance of the substrate, and $d$ is the thickness of the film. A continuous *l*-He flow cryostat was used for the temperature-dependent measurements. The dependence of the optical spectra on the magnetic field was investigated between 1.61 eV (770 nm) and 2.34 eV (530 nm) using a grating-type spectrophotometer and a 33 T resistive magnet at the National High Magnetic Field Laboratory, Tallahassee, FL.

## III. RESULTS AND DISCUSSION

Figure 1 shows the absorption spectra of the hexa-$R$MnO$_3$ thin films ($R$ = Gd, Tb, Dy and Ho) at various temperatures. For clarity, only the optical data at selected temperatures, i.e., 15 K, 100 K, 200 K, and 300 K, are shown. A general spectral feature observed in all the hexa-$R$MnO$_3$ thin films is the strong absorption structure near 2.0 eV with significant temperature dependence, accompanied by a much stronger charge transfer absorption (O 2$p$ to Mn 3$d$ states) in the higher photon energy region above 3 eV. According to our previous study,[18] this structure consists of a major peak at ~1.7 eV and a secondary peak at ~2.3 eV, which are attributable to the inter-site optical transition from the hybridized occupied states with $d_{xy}/d_{x^2-y^2}$ and the $d_{yz}/d_{zx}$ orbital symmetry to the unoccupied Mn $d_{3z^2-r^2}$ state, respectively. The existence of the second peak was supported by the asymmetric shape of the ~2.0 eV absorption structure with a larger spectral weight at higher energy. While these overall features appeared to be insensitive to the temperature variation, it is intriguing to observe additional fine structures developing in low-temperature absorption spectra exclusively for hexa-GdMnO$_3$, as indicated with arrows in Fig. 1(a). One possible explanation for the additional structure is the strong magnetic interaction between the Gd spins. In hexa-GdMnO$_3$, the interaction between $R$ ion spins is the greatest among the hexa-$R$MnO$_3$ series studied here since Gd ions have the largest magnetic moment. The large magnetic dipole interaction could lead to additional distortion of the crystal structure.[19, 20] This would reduce the crystallographic symmetry, which would allow additional optical absorptions at lower temperatures where the thermal broadening effect cannot screen the small electronic fine structures.[16, 21] The contingent presence of external defects could also be another possible explanation. Bulk $GdMnO_3$ compound is located furthest from the phase boundary of orthorhombic and hexagonal structures in the $R$MnO$_3$ series. This implies that an orthorhombic phase is relatively more stable in $GdMnO_3$ as compared to the other $R$MnO$_3$ samples studied. Therefore, as $GdMnO_3$ requires the largest amount of strain energy for stabilization into a hexagonal structure, the chances of the formation of unexpected phases such as vacancies and defects are higher, and this could also explain the additional fine structures in the optical spectra.

We now focus in further detail on the temperature dependence of the absorption structure near 1.7 eV in hexa-$R$MnO$_3$. All of the hexa-$R$MnO$_3$ thin films show a similar continuous blue shift and sharpening of the ~1.7 eV peak with decreasing temperature, while the spectral weight of the peak structure remains nearly constant. To obtain further insight, we examined the peak position of the corresponding structure as a function of temperature (Fig. 2). The peak positions for all of the hexa-$R$MnO$_3$ thin films exhibit nearly the same behavior, which indicates continuous hardening with a decreasing temperature. This behavior agrees with the results obtained previously for bulk hexa-$LuMnO_3$, the data for which are also included in Fig. 2.[13] For a systematic comparison between the hexa-$R$MnO$_3$ samples, we normalized the experimental temperature with $T_N$ ($T_{Nn} \equiv T / T_N$), where the values of 56 K, 70 K, 60 K, 75 K, and 90 K were used as $T_N$ for $R$ = Gd, Tb, Dy, Ho, and Lu, respectively.[12, 13, 15, 16, 22] The peak positions were also normalized to the peak position at $T_N$ ($\omega_n$). As shown in Fig. 2(a), the normalized spectra for all hexa-$R$MnO$_3$ compounds show consistent temperature-dependent features. The largest change in $\omega_n$ can be observed near $T_N$. However, the change in $\omega_n$ persist even at temperatures fivefold higher than $T_N$. In addition, the change in $\omega_n$ for $T > T_N$ is about 3 times larger than that for $T < T_N$. These observations suggest that all hexa-$R$MnO$_3$, including both single-crystal bulk samples and thin film samples, should experience similar changes in electronic structure in close relation with the magnetic state.

To observe the temperature dependence of the ~1.7 eV peak position more clearly, we calculated the absolute differentiated values of the normalized peak position at each temperature point. As shown in Fig. 2(b), there was a clear peak at $T_N$ in the absolute differentiated values of the normalized peak position, which indicated that the rate of the peak shift was fastest when the temperature crossed $T_N$. The largest change in the optical spectra across $T_N$ implies that a significant change in the electronic structure of hexa-$R$MnO$_3$ occurs due to the effect of long-range antiferromagnetic ordering. This behavior strongly suggests coupling between the electronic structure and the magnetic structure, mediated through a strong charge-spin interaction.

While the largest change of the optical spectra near $T_N$ indicates effective coupling between the electronic structure and the long-range magnetic ordering, it should be noted that the shift of the absorption peak near 1.7 eV over the whole measurement temperature range is still quite significant. One possible reason for this unexpectedly large peak shift may be the temperature-dependent lattice effect. Generally, as the lattice constant changes, the inter-site optical transition shifts due to the change in the difference between the Madelung potentials. In this sense, the lattice contraction usually leads to a blue shift of the inter-site optical transition. A recent neutron scattering on bulk (Y, Lu)$MnO_3$ samples indicated that the lattice volume decreased by ~0.3 % when the temperature was reduced from 300 K to 10 K.[23] While this behavior appears to be accordance with our optical results of the hardening of the ~1.7 eV absorption structure, the fairly small change in lattice volume with temperature variation makes it difficult to explain the relatively large temperature dependence of the ~1.7 eV structure, where the peak shifts in our hexa-$RMnO_3$ samples were typically 0.15–0.2 eV. We also examined the effects of local distortion of the $MnO_5$ triangular bipyramids on the electronic structure. Our previous study indicates that the increase in $a/c$ ratio should increase the crystal field splitting in Mn ions through flattening of the $MnO_5$ bipyramids, leading to a blue shift of the ~1.7 eV peak position.[18] On the basis of this reasoning, it can be speculated that the hardening of the ~1.7 eV structure with decreasing temperature could be due to flattening of the $MnO_5$ bipyramids (or an increase of the $a/c$ ratio). However, this simple explanation is contrary to the results of neutron scattering experiments where the $a/c$ ratio was shown to decrease with decreasing temperature for bulk (Y, Lu)$MnO_3$ samples.[23] Therefore, we concluded that the lattice effect can be excluded as an explanation for the temperature-dependent peak shift.

A more plausible explanation was given by Souchkov *et al.*, who suggested that the large peak shift near 1.7 eV between ~10 K and ~300 K could be attributed to the exchange energy in hexa-$LuMnO_3$.[13] This seems reasonable as the short-range antiferromagnetic interaction should remain at temperatures below $\theta_{CW}$. Indeed, the optical process occurs locally and it could be sensitive to short-range magnetic ordering. While the strong geometrical frustration in hexa-$RMnO_3$ compounds suppresses the long-range antiferromagnetic ordering down to $T_N$, the strong short-range antiferromagnetic exchange interaction could persist up to $\theta_{CW}$ ($\gg T_N$).[24] As this antiferromagnetic fluctuation is suppressed with increase in temperature, the band gap, which is strongly affected by the antiferromagnetic interaction, should be reduced and the absorption structure near 1.7 eV should shift to lower energies at higher temperature. In this context, it is expected that the temperature dependence of the ~1.7 eV structure should remain at temperatures below $\theta_{CW}$, which is couple of hundreds Kelvin. Indeed, all spectra of the hexa-$R$MnO$_3$ series show that the peak shift does not slow down at 300 K. Hence, this unexpectedly large peak shift is another strong piece of evidence for the coupling of the electronic structure to the magnetic ordering, along with the anomaly at $T_N$ in all of the hexa-$R$MnO$_3$ compounds.

To obtain further evidence of the effects of the magnetic structure on the electronic structure, we obtained optical transmittance measurements under a high external magnetic field. Previously, Rai *et al*. barely identified very small magnetodielectric effects of the ~1.7 eV peak for hexa-HoMnO$_3$ by measuring magnetic field-dependent reflectance spectra. In the present study we introduced the measurement of transmittance under a strong magnetic field. It is well known that measuring the transmittance rather than the reflectance is more suited to identifying minute changes in optical spectra. This is because the transmittance is connected to the optical constants in a more straightforward way and does not involve any numerical calculations, such as the Kramers-Kronig transformation.[25, 26] By taking advantage of the thin film geometry, we were able to obtain the transmittance optical spectra that could provide fairly accurate data of the magnetic field dependence of the ~1.7 eV peak.

We measured the optical transmittance spectra of hexa-TbMnO$_3$ thin films, while applying a magnetic field along *c*-axis (Faraday geometry) up to 32 T. Figures 3(a) and 3(b) show the transmittance spectra near 2.0 eV under various magnetic fields at 30 K (below $T_N$) and 100 K (above $T_N$), respectively. For ease of comparison, we normalized the transmittance spectra at different magnetic fields by the spectra at the zero field, i.e., $Tr$ ($H$) / $Tr$ ($H$=0 T). The results clearly indicated that deviation of the spectral ratio from unity becomes larger as the magnetic field increases. The development of the distinct two peaks with opposite signs was attributed to the shift of the ~1.7 eV peak to lower energy. The absorption for different magnetic fields is shown in Figs. 3(c) and 3(d), where a continuous red shift of the ~1.7 eV peak was identified with increasing magnetic field strength. For clarity, the data near the peak are magnified in the insets of Figs. 3(c) and 3(d).

Figures 4(a) and 4(b) show the positions of the ~1.7 eV peak as a function of the magnetic field at 30 K and 100 K, respectively. At 30 K, which is below $T_N$, the peak shift was smaller than that at 100 K, and the peak was quite stable against external magnetic fields up to around 20 T, where it began to decrease continuously in energy. On the other hand, at 100 K, which is above $T_N$, the peak appeared to be more susceptible to the external magnetic field. The peak began to shift almost immediately to a lower energy when the magnetic field was applied.

Our magnetic field results also implied that the magnetic ordering influences the electronic structure of the hexa-$RMnO_3$ in accordance with the temperature-dependent measurements. A sufficiently strong external magnetic field can disturb the long-range antiferromagnetic ordering structure. This will reduce the band gap, which is affected by the magnetic correlation,[27] and eventually decrease the peak energy of the ~1.7 eV peak. This may be supported by the simple comparison in terms of the energy scale. The 20 T magnetic field where the peak energy began to decrease at 30 K corresponds to the temperature of 28 K in the energy scale. This value is comparable to the difference between the measurement temperature and $T_N$, which suggests that an energy of around 70 K could break the long-range spin ordering. Above $T_N$, the long-range magnetic ordering has already been broken, and so the peak shifts instantly with application of an additional magnetic field. Our idea was supported by the similar slope value for 30 K above 20 T and for 100 K, which is about $-3.2\times10^{-4}$ (eV/T), where the origin of the peak shift could be considered the same.

The temperature- and magnetic field-dependent optical results manifest the coupling between the electronic and magnetic structure in multiferroic hexa-$RMnO_3$ materials. A similar analogy can also be applied to CuO, which recently has been identified as a high-temperature multiferroic material.[28] The antiferromagnetic ground state of CuO is affected by the geometrical frustration that is due to its monoclinic structure. A previous optical study indicated that the optical gap of CuO shows a strong blue shift of about 200 meV when the temperature is decreased from 300 K to 10 K. This experimental optical gap shift behavior can be explained theoretically in terms of the coupling of the electronic structure to the magnetic structure.[29] According to density functional theory calculations, the reduction of the antiferromagnetic volume fraction pushes the optical gap to a lower energy with increasing temperature, similar to what is seen in hexa-$RMnO_3$ series. While the energy scale in temperature dependence is comparable to the case of the hexa-$RMnO_3$ series, a distinct anomaly is not observed at the magnetic phase transition temperatures. Instead, the gradient of the optical absorption peak changes near $T_N$, implying that the magnetic ordering could alter the character of the optical gap. Together with the results from the hexa-$RMnO_3$ compounds, this strong dependence of the electronic response on the magnetic ordering identified in CuO may indicate that the coupling between the electronic structure and magnetic structure is common in multiferroic oxide system.

## IV. SUMMARY

In summary, we have carried out an extensive optical investigation on the multiferroic hexagonal $RMnO_3$ series, and determined the general behavior of its electronic structure in view of charge-spin coupling. Using temperature-dependent transmission spectroscopy, we studied the behavior of the lowest lying inter-site optical absorption peak near 1.7 eV. The peak showed an unexpected large shift over the whole measurement temperature, accompanying the maximum shift near the Néel temperature. This behavior could be properly attributed to the effect of antiferromagnetic correlation on the electronic structure. These absorption spectra, which are related to the charge-spin coupling, also exhibit sizable magnetic field dependence. Both our temperature- and magnetic field-dependent measurements indicate strong coupling between the magnetic ordering and the electronic structure in the multiferroic hexagonal $RMnO_3$ compounds.

ACKNOWLEDGMENTS

We are grateful to H. D. Drew, J. G. Park, J. Yu, D. G. Kim, and Y.-D. Jho for valuable discussions. This study was financially supported by Creative Research Initiatives (Functionally Integrated Oxide Heterostructures) of the Ministry of Science and Technology (MOST) and the Korean Science and Engineering Foundation (KOSEF). Y.S.L. was supported by the Korea Research Foundation Grant funded by the Korean Government (MOEHRD, Basic Research Promotion Fund) (KRF-2007-314-c00088) and the Soongsil University Research Fund. A portion of this work was performed at the National High Magnetic Field Laboratory, which is supported by NSF Cooperative Agreement No. DMR-0084173, by the State of Florida, and by the DOE.

**FIGURE CAPTIONS**

Fig. 1. (Color online) Temperature-dependent in-plane optical spectra of (a) $GdMnO_3$, (b) $TbMnO_3$, (c) $DyMnO_3$, and (d) $HoMnO_3$ thin films. The arrows in (a) indicate additional fine structures seen at low temperature.

Fig. 2. (Color online) Temperature dependence of the optical transition at ~1.7 eV. (a) Temperature dependence of the peak positions normalized to the peak position at $T_N$ ($\omega_n$) as a function of the normalized temperature $T_{Nn}$ ($\equiv T / T_N$). Temperature values of 56, 70, 60, 75, and 90 K were used as $T_N$ and the energy values of 1.9725, 1.77, 1.785, 1.7711, and 1.7249 eV were used as the peak energy values at $T_N$ for $R$ = Gd, Tb, Dy, Ho, and Lu, respectively. (b) Absolute value of normalized peak position differentiated to normalized temperature.

Fig. 3. (Color online) Magnetic field dependence of the optical transition at ~1.7 eV of hexa-$TbMnO_3$ thin film (a)(c) at 30 K and (b)(d) at 100 K. (a)(b) Transmittance ratio, $Tr(H) / Tr(H$=0 T) with magnetic field along the *c*-axis from 0 to 30 T. The data are shown in 5 T steps. (c)(d) Absorption spectra illustrating actual peak shifts. The insets show the enlarged area of the gray rectangular in (c) and (d).

Fig. 4. The peak position of the optical transition at ~ 1.7 eV plotted as a function of magnetic field at (a) below $T_N$ (30 K) and (b) above $T_N$ (100 K). The arrow in (a) indicates the starting point of the peak shift. The gray linear lines are guides to eyes. The photon energy scales of the *y*-axes are the same for direct comparison.

* Electronic address: ylee@ssu.ac.kr

[a] Present address: Max-Planck-Institut für Festkörperforschung, Heisenbergstr. 1, D-70569 Stuttgart, Germany

[b] Present address: Advanced Technology Development Team, Samsung Electronics, Gyeonggi-Do, 449-711, Korea

[c] Present address: Department of physics, Indian Institute of Technology Madras, Chennai 600036, India

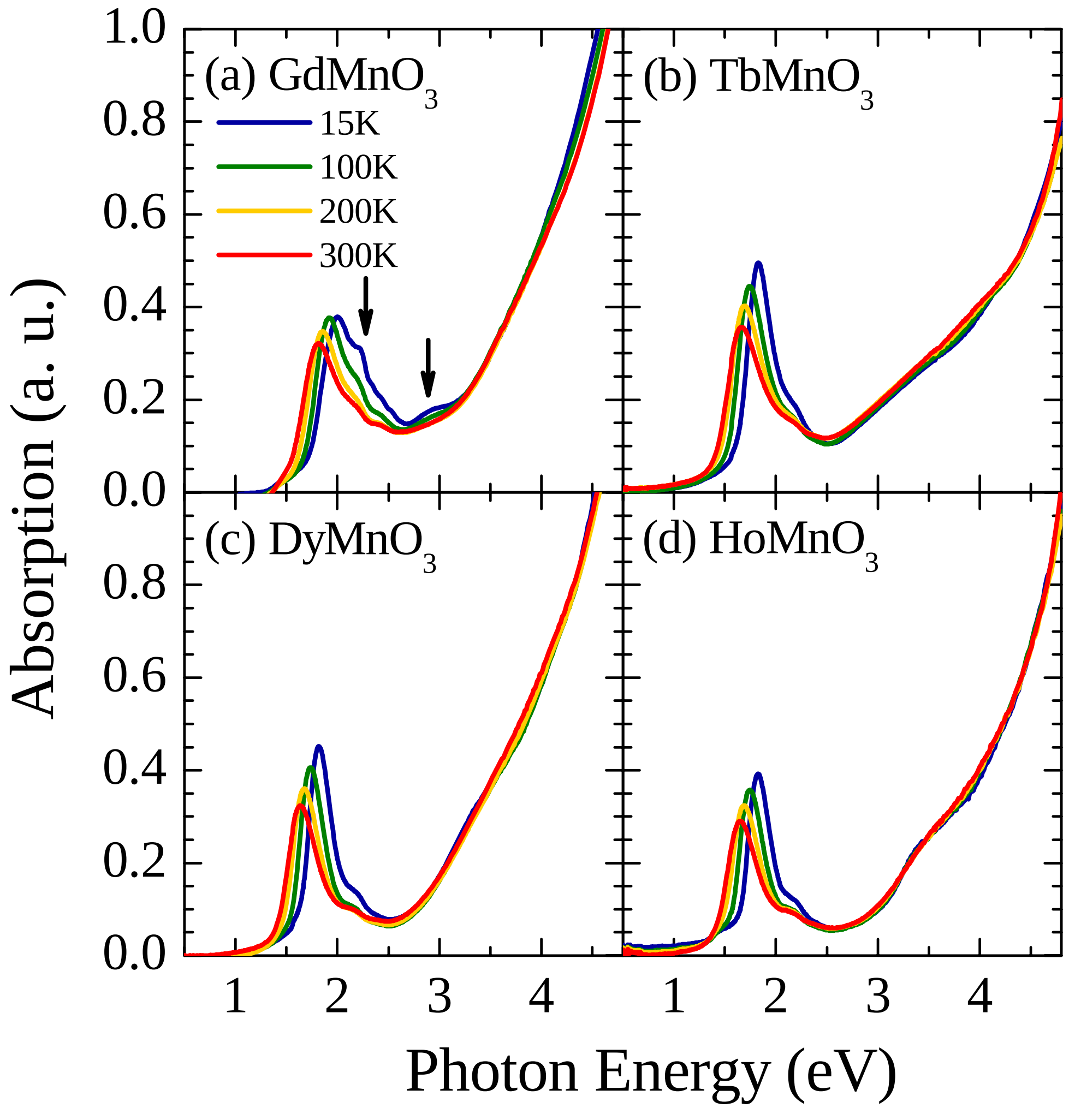

Fig. 1.

Choi *et al.*

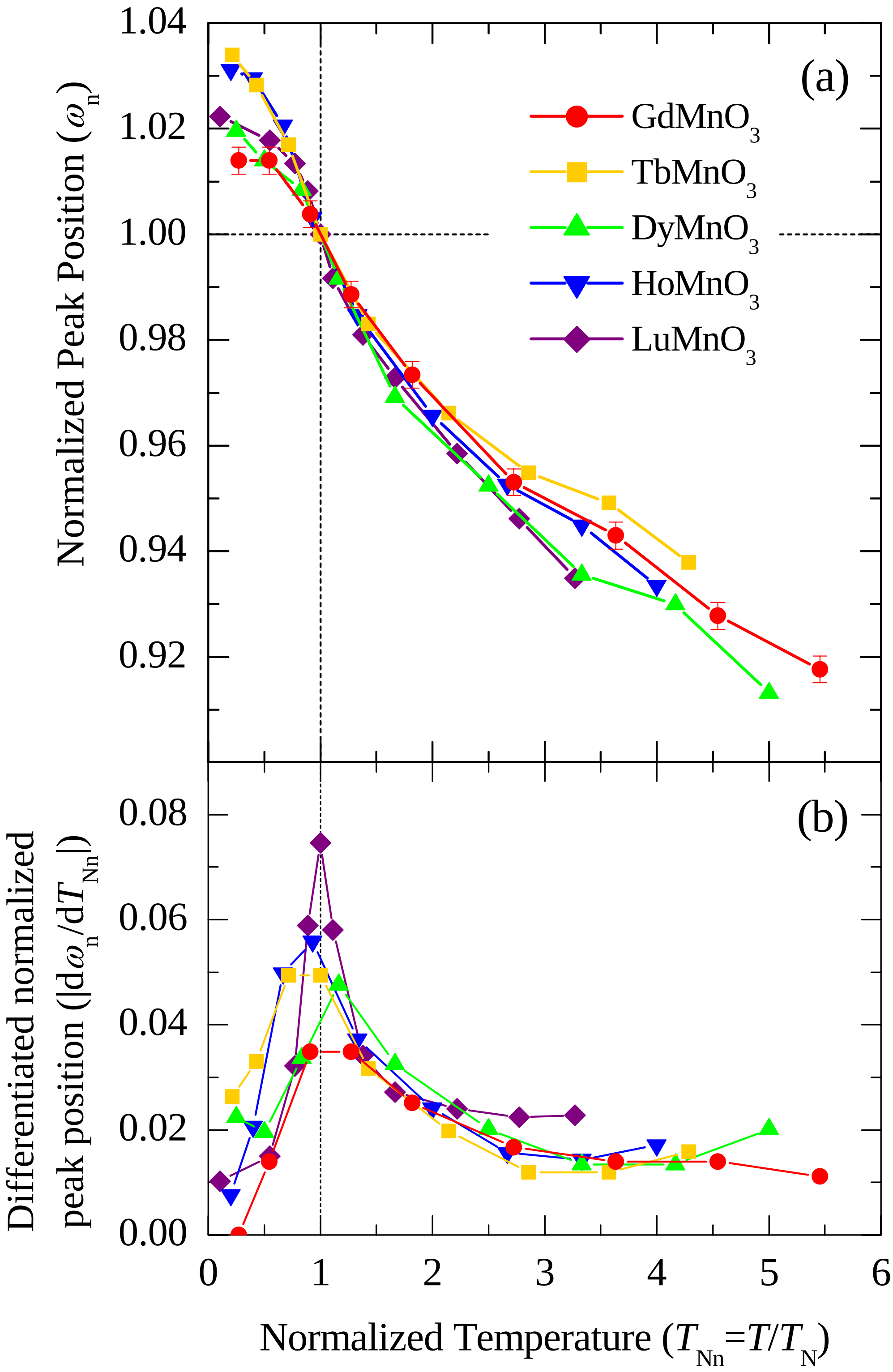

Fig. 2.

Choi *et al.*

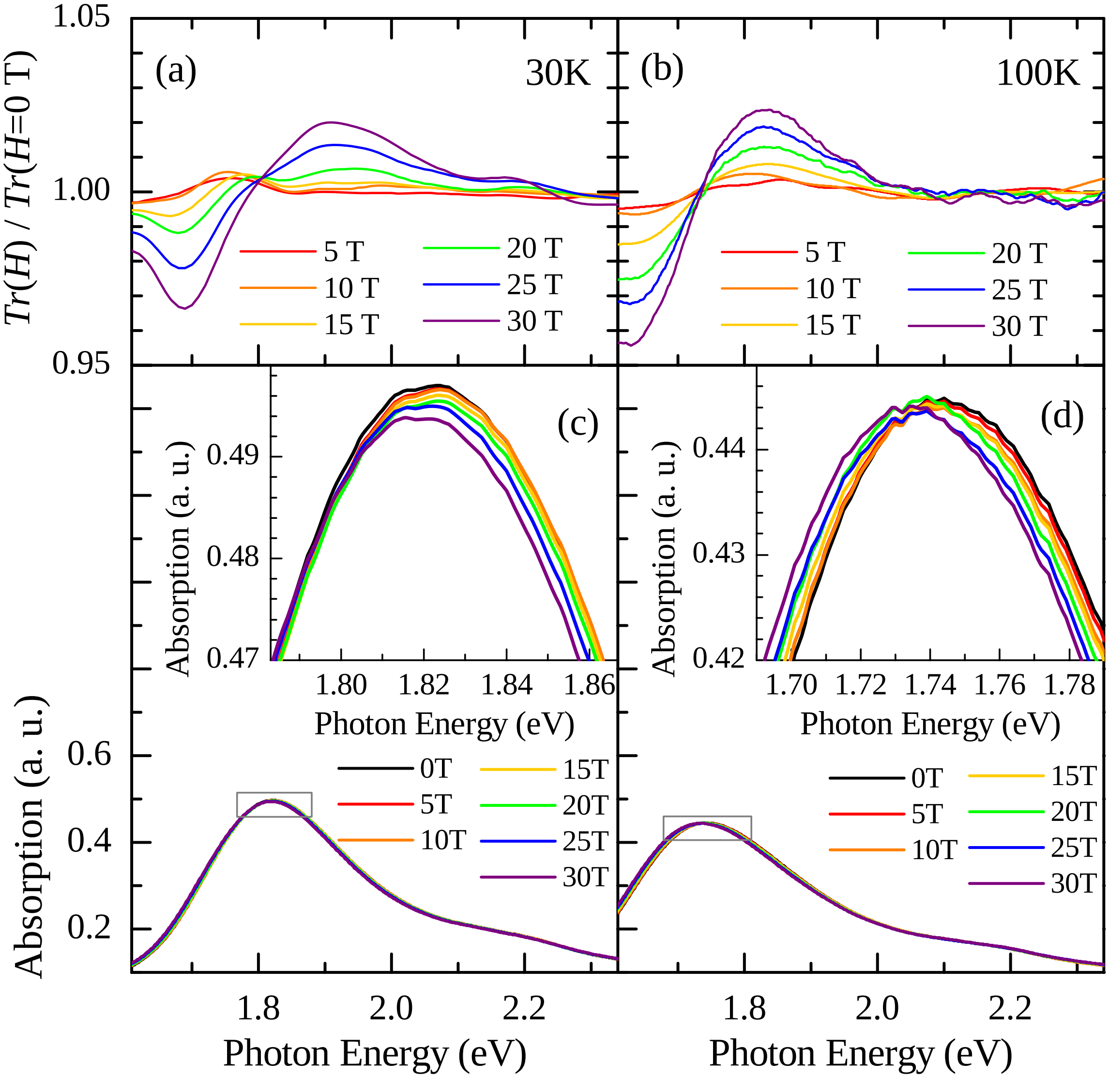

Fig. 3.

Choi *et al.*

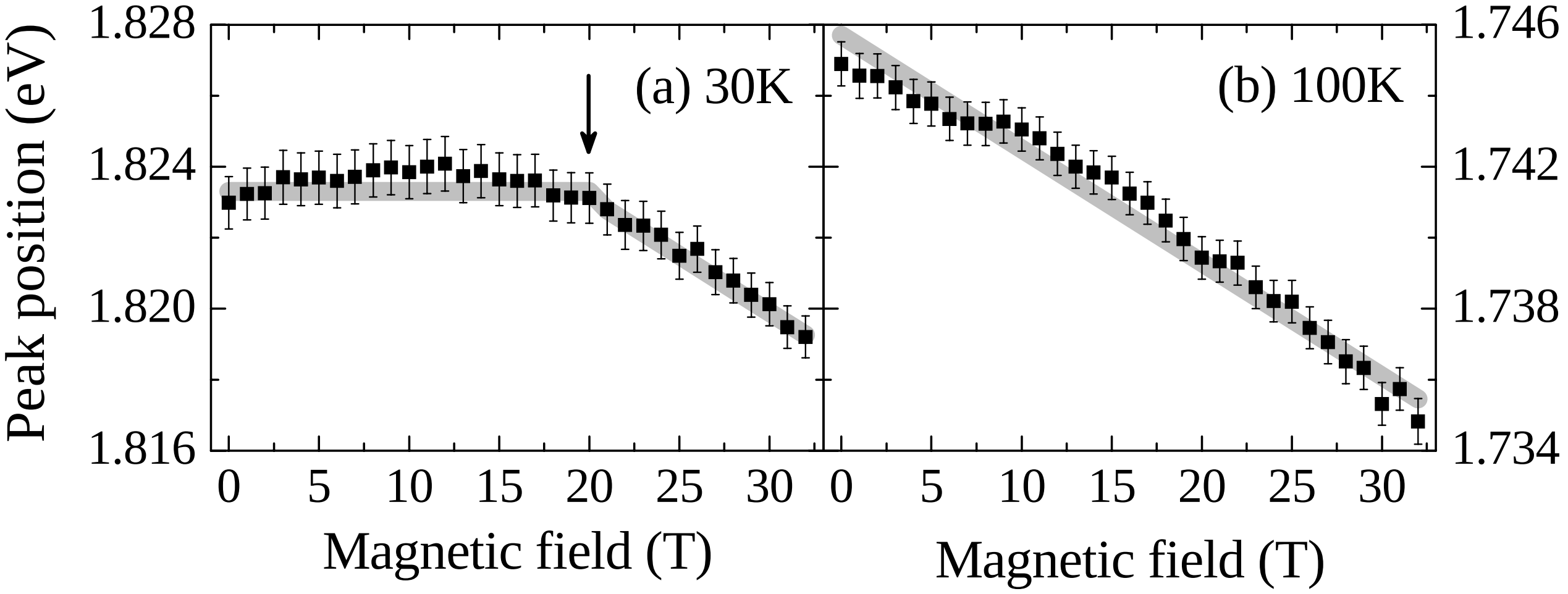

Fig. 4.

Choi *et al.*